\documentclass[a4paper,11pt]{article}
\usepackage{pos}
\usepackage{hyperref}

\newcommand{\equ}[1]{Eq.~(\ref{equ:#1})}
\newcommand{\figu}[1]{Fig.~\ref{fig:#1}}

\title{Sources of high-energy astrophysical neutrinos}

\author*{Walter Winter}

\affiliation{Deutsches Elektronen-Synchrotron DESY, \\
Platanenallee 6, 15738 Zeuthen, Germany}

\emailAdd{walter.winter@desy.de}

\abstract{We discuss recent results in neutrino astronomy and their implications for the cosmic-ray acceleration in relativistic outflows, such as in Active Galactic Nuclei (AGN) jets, Gamma-Ray Bursts (GRBs), and Tidal Disruption Events (TDEs). We especially focus on challenges at the interface to particle acceleration which can be inferred from the multi-messenger context, such as the paradigm that the sources power the Ultra-High-Energy Cosmic Rays (UHECRs). We demonstrate that both AGN blazars (in the context of neutrino observations) and GRBs (as UHECR sources in the context of neutrino-non-observations) point towards  acceleration spectra harder than $E^{-2}$, or relatively high minimal cosmic-ray injection energies, to meet the respective energy budget requirements. We furthermore speculate that neutrino flares in blazars may be related to super-Eddington accretion flares, or that GRBs are powered by significantly higher kinetic energies than typically assumed in electromagnetic models. For internal shock models, the UHECR paradigm for GRBs can only be maintained in the light of neutrino stacking limits in multi-zone models. While relativistic outflows in TDEs have become recently interesting {\em per se} and models for the neutrino emission from jetted TDEs exist, a direct connection between TDE jets pointing in our direction and astrophysical neutrinos has not been identified yet.}

\FullConference{High Energy Phenomena in Relativistic Outflows VIII (HEPROVIII)\\
 23-26 October, 2023\\
Paris, France\\}

\begin{document}
\maketitle

\section{Introduction}

Since a diffuse astrophysical neutrino flux has been discovered by the IceCube experiment~\cite{IceCube:2013low}, several individual source contributions have been identified: neutrinos from the Active Galactic Nuclei (AGN) blazar TXS 0506+056~\cite{IceCube:2018dnn,IceCube:2018cha}, neutrinos from the active galaxy NGC 1068~\cite{IceCube:2022der}, and neutrinos from the Galactic plane~\cite{IceCube:2023ame}. In addition, there have been a number of interesting constraints on source classes which are believed to contribute significantly to the astrophysical flux. For example, Gamma-Ray Bursts (GRBs) probably contribute less than about one percent of the diffuse flux~\cite{IceCube:2012qza}. Apart from the well established neutrino sources, a number of additional contributors with some hints for neutrino-source associations have been proposed; one example are Tidal Disruption Events (TDEs)~\cite{Stein:2020xhk,Reusch:2021ztx,vanVelzen:2021zsm,Yuan:2024foi}.

Relativistic outflows are typically associated with particle acceleration leading to non-thermal primary spectra. Radiative mechanisms produce secondary electromagnetic signatures, such as synchrotron emission of electrons and inverse Compton up-scatterings of those photons by the same electrons (synchroton self-Compton models). If protons (or nuclei) are acclelerated as well, neutrinos may be produced by the interactions of protons and photons or matter, as well as additional (so-called ``hadronic'') signatures on the electromagnetic spectrum are expected. One of the outstanding puzzles in astroparticle physics in the multi-messenger context is the origin of the Ultra-High-Energy Cosmic Rays (UHECRs), which are the cosmic rays at the highest energies. More specifically, if one expects significant neutrino production in the sources of the cosmic rays, astrophysical neutrinos will be a smoking gun signature for their origin.

Here we focus (in the spirit of this conference series) on source classes which potentially host relativistic outflows, and where the neutrino production might be associated with those outflows.  A widely accepted neutrino source class are AGN jets, where the Lorentz factor of the outflow  $\Gamma \sim 10-30$. AGN jets are best studied as AGN blazars, which are jets pointing in our direction, in terms of data coverage across the spectrum; electromagnetic data typically give a lot of information on the properties of the emission region and constrain its parameters. Note, however, that because of the isotropy of the Universe, we expect that AGN jets pointing in other directions have (statistically) the same properties as AGN blazars, even though they may be classified into different categories from the observational point of view. Another very interesting source class are GRBs, where we focus on the prompt emission of standard long-duration GRBs with $\Gamma \gtrsim 100$ in the context of internal shock models for the sake of simplicity. GRBs are one of the main candidate classes for the origin of the UHECRs because of their energetics, and they are also expected to produce astrophysical neutrinos~\cite{Waxman:1997ti}. The non-observation of neutrinos from GRBs~\cite{IceCube:2012qza} therefore has  profound consequences for the UHECR paradigm for GRBs, as we will discuss. It has been also well established that some TDEs come with relativistic jets since the discovery of of Swift
J1644+57~\cite{Burrows:2011dn}; while several likely observations of TDEs with jets have been made since then, the subject has recently gained momentum by the observation of AT2022cmc \cite{Andreoni:2022afu}; the inferred $\Gamma \sim 10-100$~\cite{Burrows:2011dn,Pasham:2022oee,Yao:2023jtk}. The association between TDE jets and astrophysical neutrinos is somewhat more speculative, as we will see.

The purpose of this talk is {\em not} to disucss how the particles are accelerated, but give ideas what it needs to describe the astrophysical neutrino observations in the multi-messenger context --  and what the critical issues are. For example: What kind of acceleration spectra do we need? What do we expect for the energy budget? What injection composition do we need to describe UHECRs?  

\section{Physics of neutrino production in the multi-messenger context}
\label{sec:mm}

Assuming proton primaries, the neutrino production in relativistic outflows is frequently dominated by $p\gamma$ interactions. These are frequently approximated by the $\Delta$-resonance for the sake of simplicity:
\begin{equation}
p + \gamma \rightarrow \Delta^+ \rightarrow \left\{ \begin{array}{lll} 
n + \pi^+  & \mathrm{branching} \, \, \frac{1}{3} & \rightarrow \nu \\ 
p + \pi^0  & \mathrm{branching} \, \,  \frac{2}{3} & \rightarrow \gamma
\end{array}  \right. \, . \label{equ:nuprod}
\end{equation}
The first row corresponds to the production of three neutrinos and one positron from the pion decay chain, the second row to two the production of two gamma-rays. The neutrinos will carry about $E_\nu \simeq 0.05 \, E_p$, whereas the gamma-rays have about $E_\gamma \simeq 0.1 \, E_p$, i.e., about 2~PeV $\gamma$-rays correspond to a PeV neutrino. Neutrinos and gamma-rays therefore follow the primary energy and are co-produced at similar rates, but the gamma-rays are frequently re-processed within the source. Taking the pitch angle-averaging between proton and photon into account, the relevant target photon energy is typically in the X-ray range ({\em cf.}, Fig. 4a in \cite{Hummer:2010vx}):
\begin{equation}
 E_\gamma^{\mathrm{target}} \, \mathrm{[keV]} \simeq \left( \frac{\Gamma}{10} \right)^2 \,  \frac{1}{E_\nu \, \mathrm{[PeV]}} \, . \label{equ:target}
\end{equation}
For power-law proton $dN_p/dE_p \propto E_p^{-\alpha}$ and photon $dn_\gamma/d \varepsilon_\gamma \propto \varepsilon^{-\beta}$ spectra, the neutrino spectrum follows $dN_\nu/dE_\nu \propto E_\nu^{-\alpha+\beta-1}$ since high-energy protons interact with low energy photons to match the right center-of-mass energy of the $\Delta$-resonance. As a consequence, the typical neutrino spectrum $E_\nu^2 dN_\nu/dE_\nu$ for $\alpha \simeq 2$ and $\beta > 1$ increases up to the peak energy determined by $E_{\nu,\mathrm{max}} \simeq 0.05 \, E_{p,\mathrm{max}}$, such as for AGN blazars. For GRBs, the spectrum can be flat in $E_\nu^2 dN_\nu/dE_\nu$ if $\beta \simeq 1$ below the photon break (high-energy neutrinos come from low-energy target photons), and the maximal neutrino energy (cutoff) can be driven by the efficient synchrotron cooling of the secondary muons and pions in magnetic fields (see e.g. App.~A in \cite{Baerwald:2011ee} for a detailed example). 

As far as the neutrino luminosity (per flavor) $L_\nu$ is concerned, it is frequently related to the gamma-ray luminosity $L_\gamma$ via the (non-thermal) proton luminosity $L_p$ by\footnote{The formula holds in the optically thin regime, where the pion takes about 20\% of the proton energy per interaction, charged pions are (over all interaction types) produced in about 50\% of all cases, and the pion decays into four leptons.}
\begin{equation}
L_\nu \simeq  \frac{1}{8} \, 0.2 \,  L_p \, \tau_{p\gamma} = 0.025 \,  \frac{L_\gamma}{f_e} \, \tau_{p\gamma} \, ,
\label{equ:energy}
\end{equation}
where $\tau_{p\gamma}$ is the optical thickness to $p \gamma$ interactions (sometimes related to the ``pion production efficiency'' $f_\pi \simeq 0.2 \, \tau_{p\gamma}$), and $f_e^{-1}$ is the ``baryonic loading'' defined as ratio between luminosity in non-thermal protons $L_p$ to electrons $L_e$ (and assuming that electrons and photons are in equipartition $L_\gamma \simeq L_e$). Apparently two factors control the neutrino production: the baryonic loading and the optical thickness $\tau_{p\gamma} \sim R/\lambda_{\mathrm{mfp}}$ related to source size $R$ and mean-free path $\lambda_{\mathrm{mfp}}$ of the $p\gamma$ interactions, and thus the density of the target photons (or the compactness of the source). Here very big differences among the source classes emerge: while AGN blazars prefer extremely large $f_e^{-1}$ and small $\tau_{p\gamma}$ in terms of the parameters needed to describe the electromagnetic spectrum and neutrinos, GRB and TDE jets rather point towards smaller (but nevertheless beyond equipartition) $f_e^{-1}$ with more efficient $\tau_{p\gamma}$. Note that the baryonic loading $f_e^{-1} = L_p/L_e$ is frequently constrained by the overall luminosity (or energy) which can be extracted from the systems, such as in terms of the Eddington luminosity (AGN), the kinetic luminosity of the jet (GRB), or the mass fallback rate of the disrupted star (TDE), which means that one would generically only expect a small fraction of these luminosities ending up in non-thermal protons.

\begin{figure}[t]
\begin{center}
\includegraphics[width=0.4\textwidth]{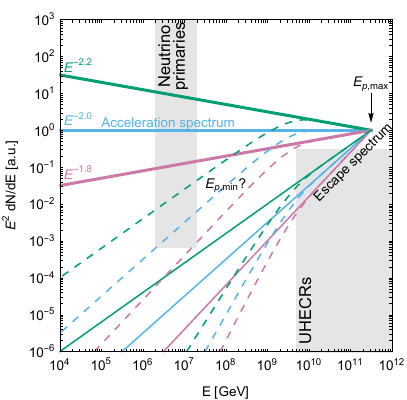}
\end{center}
\caption{\label{fig:pedplot} Illustration of the acceleration (in source) spectrum for different power laws, compared to the escape spectrum for Bohm-like diffusion. The shaded regions show the cosmic-ray energy ranges relevant for neutrino production (0.1-1~PeV) and UHECRs, the dashed curves illustrate the effect of a low-energy cutoff (here at $E_{p,\mathrm{min}} \simeq 5 \, 10^{9} \, \mathrm{GeV}$) on the cosmic-ray energy. Assumptions: Optically thin case (to any kind of interactions), Larmor radius $R_L \simeq R$ at the highest energies, and proton primaries (figure original).  }
\end{figure}

In order to describe UHECR data, a certain luminosity $L_{\mathrm{UHECR}}$ in the respective energy range is needed, as we will discuss later. For protons, this range is close to  $E_{p,\mathrm{max}}$, as illustrated in \figu{pedplot}. In most cases, it assumed that a power law proton spectrum with index $\alpha \simeq 2$ extends  from $E_{p,\mathrm{min}} \simeq 1 \, \mathrm{GeV}$ (or $\Gamma \times 1 \, \mathrm{GeV}$) to $E_{p,\mathrm{max}}$ (thick solid lines). It is easy to show that the required total total energy or luminosity scales with 
\begin{equation}
 L_p \propto \frac{1}{T} \, \int\limits_{E_{p,\mathrm{min}}}^{E_{p,\mathrm{max}}} E_p \, \frac{dN_p}{dE_p} \, dE_p \propto \left\{  
\begin{array}{ll}
 \log \left( \frac{E_{p,\mathrm{max}}}{E_{p,\mathrm{min}}} \right) & \mathrm{for} \, \, \alpha=2 \\
 ( E_{p,\mathrm{max}})^{(2 - \alpha)} & \mathrm{for} \, \, \alpha < 2  \quad . \\
 ( E_{p,\mathrm{min}})^{(\alpha -2 )} & \mathrm{for} \, \, \alpha > 2 \\
\end{array}
\right.  
\label{equ:energybudget}
\end{equation}
Thus, for $\alpha=2$, the overall energy budget only scales logarithmically with minimal and maximal energies, but for softer spectra, the minimal energy determines the energy budget, and for harder spectra, the maximal energy. 

Especially if one normalizes the spectrum to the UHECRs, as illustrated in \figu{pedplot}, the question how $L_p$ (total luminosity across the spectrum for protons or nuclei) relates to $L_{\mathrm{UHECR}}$ arises (often defined in the energy range beyond $5 \cdot 10^9 \, \mathrm{GeV}$). This may be quantified by a ``bolometric correction factor'' $f_{\mathrm{bol}} \equiv L_{\mathrm{CR}}/L_{\mathrm{UHECR}}$, which can be obtained from similar considerations to \equ{energybudget}. For the (required) energy budget, therefore not only the maximal cosmic ray-energy matters, but also the minimal $E_{p,\mathrm{min}}$ and the spectral shape. For $\alpha=2$, $f_{\mathrm{bol}} \simeq 6-10$, for $\alpha < 2$, $f_{\mathrm{bol}} \simeq 1$ (because the energy budget is driven by the maximal energy), and for the more frequently inferred acceleration spectra $\alpha = 2.2$ ($2.5$), $f_{\mathrm{bol}} \simeq 30-100$ ($4000-71000$) (because the energy budget is driven by the minimal energy). It is therefore typically difficult to accommodate acceleration spectra with $\alpha \gtrsim 2$ with the overall energy budget required to power the UHECRs unless the accelerator receives ``pre-accelerated'' cosmic rays beyond $E_{p,\mathrm{min}}$ -- similar to $E_{e,\mathrm{min}}$ frequently used in AGN blazars models. We will come back to this ``energy'' challenge later.

We also illustrate the connection between in-source acceleration spectrum and escape spectra for different power laws with a potentially high minimal proton energy cutoff $E_{p,\mathrm{min}}$ in \figu{pedplot} for different power law acceleration spectra. Here the shaded areas show the energy ranges relevant for the neutrino and UHECR production. Since the UHECRs may be magnetically confined in the source, only the highest energetic ones may leave the source where the Larmor radius becomes comparable to the source size (unless some kind of advection mechanism is effective); while details depend on the escape model, these mechanisms typically lead to harder escape compared to acceleration spectra (see e.g. \cite{Baerwald:2013pu,Globus:2014fka,Unger:2015laa}), which we illustrate in  \figu{pedplot} as well. Naturally, there will be another correction factor relating the in-source and escaping UHECR spectra (c.f., thick versus thin curves in the figure). The figure illustrates three important points:
\begin{enumerate}
\item
 The UHECR escape spectrum at the highest energies only weakly depends on the acceleration spectral shape, the minimal energy cutoff, or even the diffusion coefficient/escape mechanism (not shown) because of the relatively narrow relevant energy range and the spectral hardening; its energy budget is determined by its maximal energy $E_{p,\mathrm{max}}$ ({\em cf.}, \equ{energybudget}).
\item
 The total  energy injected into cosmic rays strongly scales with spectral shape and minimal proton energy (for $\alpha>2$), as discussed earlier.
 \item
 The neutrino production, which is driven by the in-source cosmic-ray density, depends on the cosmic-ray spectrum at  energies significantly below the ankle, {\em i.e.}, its spectral shape and minimal energy, if the cosmic rays are normalized in the UHECR range.
\end{enumerate}
Neutrino observations are thus critical to constrain what happens at lower cosmic-ray energies. For example, consider a bold proposal: perhaps the reason why no neutrinos from GRBs have been seen is that $E_{\mathrm{CR},\mathrm{min}}$ is above the critical energy for pion cooling?

\section{Active Galactic Nuclei (AGN) blazars}

\begin{figure}[t]
\begin{center}
\includegraphics[width=0.6\textwidth]{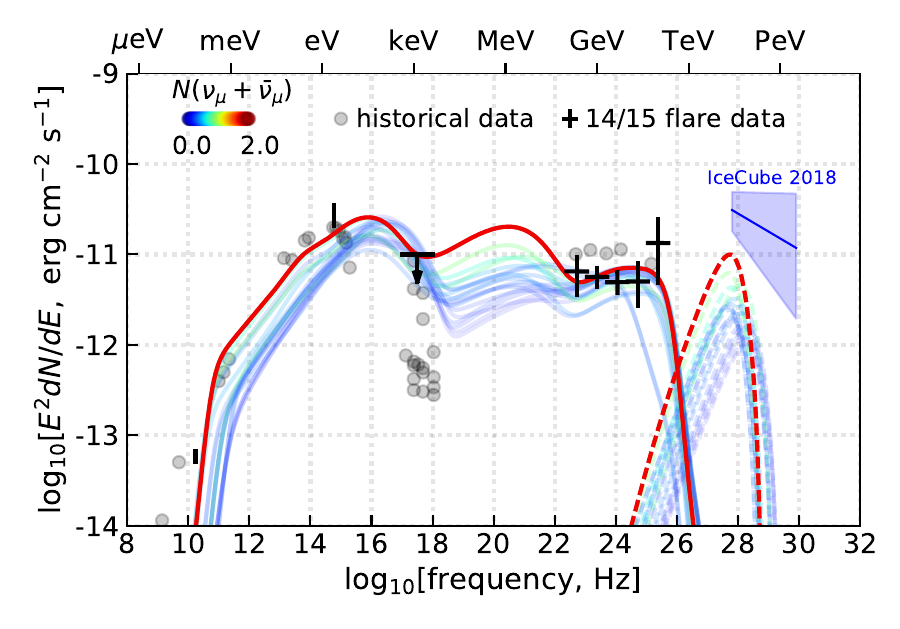}
\end{center}
\caption{\label{fig:txs} Different versions of spectral energy distributions (SEDs) and muon neutrino fluxes predicted by the one-zone hadronic model for the 2014-15 neutrino flare of TXS 0506+056, compared to the single-flavor flux derived by \cite{IceCube:2018cha}. Black data (or limits) are taken contemporously with the neutrino flare, whereas historical data are shown as gray dots.
Figure taken from \cite{Rodrigues:2018tku} (under \href{https://creativecommons.org/licenses/by/3.0/}{CC BY 3.0}).
 }
\end{figure}

The spectral energy distributions (SED) of blazars typically exhibit a double-hump structure which is believed to come from synchroton radiation of accelerated electrons and inverse Compton-upscatterings of these (or external) photons to higher energies by the same electrons.  Neutrino emission, as it has been observed in TXS 0506+056~\cite{IceCube:2018dnn,IceCube:2018cha}, however requires non-thermal protons to facilitate neutrino production, see \equ{nuprod}. Since the SED of blazars is frequently measured in multiple energy bands, often good constraints on the emitting region -- such as its size and magnetic field, and the properties of the non-thermal electron spectrum -- can be inferred. Adding the minimal ingredients for neutrino production in terms of a power law proton spectrum with spectral index $\alpha$ a maximal energy cutoff $E_{p,\mathrm{max}}$ has been only admissible as a perturbation of this picture, see e.g. \cite{Gao:2018mnu,Keivani:2018rnh,Cerruti:2018tmc,MAGIC:2018sak}. The reason lies in \equ{nuprod}: neutrinos and gamma-rays are co-produced together; while neutrinos can easily escape, the gamma-rays are typically cascaded down within the source, leading to signatures in the X-ray bands where the intrinsic leptonic emission is low (the ``dip''). These (and the very-high-energy gamma-ray) energy ranges consequently limit the baryonic loading.

While one neutrino event associated with an blazar can be explained in terms of a larger population of sources, each emitting only a very low neutrino fluence on average, the 2014-15 neutrino flare of TXS 0506+056 yielding 13$\pm$5 neutrino events has been especially  puzzling. In this case, however, no contemporary data in the X-ray range were available, which means that the re-processed gamma-ray energy may show up exactly there. This kind of hadronic model is shown in \figu{txs} for several model parameter sets (see also \cite{Reimer:2018vvw}), where an additional hump in the keV-MeV range (from hadronic processes) is visible. While this hypothesis seems quite exotic at a first glance, it highlights the importance of contemporary source monitoring across the whole electromagnetic spectrum for these kind of multi-messenger events. 

Another typical challenge for AGN blazar models with neutrino production is the low $\tau_{p \gamma} \ll 1$, which can be (in one zone models) derived from the properties of the source (such as production region size and luminosity), which needs to be paired with large $1/f_e \gtrsim 10^4$ (example for TXS 0506+056 in \cite{Gao:2018mnu}) to match a reasonable neutrino event rate. This challenge is frequently mitigated by multi-zone approaches, such as external target photons adding to the target or more compact neutrino production zones. Another possibility is to use higher $E_{p,\mathrm{max}}$ in mismatch with  the observed $E_{\nu,\mathrm{max}}$, as the target photon (number) density is higher at lower energies. Overall, it is difficult to avoid this tension between  $\tau_{p \gamma}$  and $1/f_e$ at all, which means the obtained $1/f_e$ is typically far away from energy-equipartition between electrons and protons.

So what have we learned about the neutrino emission from AGN blazars since TXS 0506+056? While numerous sources have been investigated and systematic source class-neutrino associations have been performed (see e.g. \cite{Oikonomou:2019djc,Plavin:2020emb,Buson:2022fyf}), we highlight the results from a recent study in relationship to the questions raised earlier \cite{2024A&A...681A.119R}. In that study, 324 AGN blazars have been modeled individually, which is the largest sample to date, in order to come derive statistical properties. It has been demonstrated from the Spectral Energy Distribution (SED) perspective that about 66\% of these blazars are best described with a purelty leptonic SED, about 15\% benefit from hadronic contributions in the electromagnetic spectrum, and about 20\% prefer a hadronic contribution such as to describe X-ray data better. While the required physical luminosity into non-thermal electrons $L_e^{\mathrm{phys}} \simeq (10^{-3} - 10^{-2}) \, L_{\mathrm{Edd}}$, the required physical luminosity into non-thermal protons $L_p^{\mathrm{phys}} \simeq (10^{-1} - 10^{2}) \, L_{\mathrm{Edd}}$ -- which raises the question of the neutrino emission is perhaps associated with super-Eddington accretion phases.\footnote{In this example, hard acceleration spectra and high minimal energies have been chosen already to mitigate the energy challenge.} It has been also demonstrated that the baryonic loading scales with $L_\gamma^{-0.6}$, which means that more luminuous sources seem to carry a smaller fraction of energy in non-thermal protons. While this anti-correlation has been derived here from the electromagnetic SED of different AGN blazars, qualitatively similar conclusions have been found in different contexts strengthening the case: a) The multi-Epoch Modeling of TXS 0506+056 itself exhibits a similar tendency~\cite{Petropoulou:2019zqp}. b)  It can be more fundamentally motivated in magnetically dominanted, baryon-loaded jets~\cite{Petropoulou:2022sct}. c) Postulating that the diffuse neutrino flux is dominated by AGN blazars (at the highest energies) requires strong evolution of the baryonic loading, because stacking searches limit the most luminuous source contributions to neutrinos~\cite{Palladino:2018lov}.\footnote{Note that, depending on the scaling of neutrino with photon luminosity, neutrino multiplet limits can constrain the neutrino emission per AGN blazar~\cite{Yuan:2019ucv}.}

So what do we finally learn from the neutrino emission about the accleration of protons in AGN blazars? While the electron spectrum is typically modeled with parameters such as $E_{e,\mathrm{min}}$ (or the fraction of accelerated electrons), $E_{e,\mathrm{max}}$ and spectral index $\alpha_e$, similar assumptions for protons are more delicate in terms of the overall energy budget because of the relatively required $f_e^{-1}$. This energy crisis can be overcome with relatively high $E_{p,\mathrm{min}}$ (contrary to what is typically assumed in the models) or hard acceleration spectra $\alpha_p < 2$, as we already indicated in Sec.~\ref{sec:mm}. Since the theoretical motivation for the electron spectral parameters is frequently uncertain as well, the question if the proton and electron parameters could be somehow related arises, or what fundamentally motivates these parameters (see e.g. \cite{Zech:2021emy}). In addition, it is a question what drives the large baryonic loading of neutrino-emitting AGN blazars significantly beyond energy equi-partition, and why the baryonic loading decreases with luminosity. Finally, the speculation if neutrino-emitting AGN (flares) may be related to super-Eddington accretion (perhaps similar to TDEs) arises.

\section{Gamma-Ray Bursts (GRBs)}

Since GRBs are one of the most prominent candidate classes for the UHECR origin, they are a good test case to illustrate the role of astrophysical neutrino measurements in the context of different (internal shock) GRB models, including the implications for energetics and particle acceleration. A back-of-the-enevelope estimate of the required (isotropic-equivalent) energy output (escaping UHECRs) per GRB beyond $10^{10} \, \mathrm{GeV}$ is $E^{\mathrm{esc}}_{\mathrm{UHECR}} \simeq 10^{53} \, \mathrm{erg}$ for a local (apparent) GRB rate of $1 \, \mathrm{Gpc}^{-3} \, \mathrm{yr}^{-1}$ ($10^{-9} \, \mathrm{Mpc}^{-3} \, \mathrm{yr}^{-1}$), which leads to the typical energy injection rate of $\dot \varepsilon \simeq 10^{44} \, \mathrm{Mpc}^{-3} \, \mathrm{yr}^{-1}$ roughly needed to describe UHECR data~\cite{Jiang:2020arb}. Thus, from the discussion in  Sec.~\ref{sec:mm}, it is easy to see that for  $f_e^{-1} \simeq 10$ and $f_{\mathrm{bol}} \simeq 10$ (expected for $E^{-2}$ acceleration spectra) $E^{\mathrm{esc}}_{\mathrm{UHECR}} \simeq E_\gamma^{\mathrm{iso}}$ if all UHECRs can escape, which has lead to the generic estimate of a ``baryonic loading 10'' if GRBs are to power the UHECRs. There are a number of assumptions and omitted factors though (e.g. what fraction of UHECRs can actually escape?), so that one needs to test this assumption in specific source-propagation models.

One big challenge arises from neutrino data: a baryonic loading of 10 can be excluded for typical values for the GRB luminosity and $\Gamma \simeq 300$ from stacking limits~\cite{IceCube:2012qza,IceCube:2017amx}. Even if the baryonic loading is left as a free parameter and UHECRs are fitted, it is quite safe to say under various assumptions that these typical values lead to too large neutrino fluxes in the one-zone model, i.e., if neutrinos and UHECRs come from the same collision radii $R_C$; possible ways out may be large collision radii (pointing towards magnetic reconnection models) or low luminosities (pointing towards low-luminosity GRBs)~\cite{Biehl:2017zlw}. Is therefore the UHECR paradigm for GRBs ruled out already?

First of all, it is important to note that typical internal shock models do not predict a single $R_C$, but a range of radii. This is, in fact, a necessary implication for the efficient energy dissipation (into non-thermal particles), which is proportional to the difference of the Lorentz factors of colliding plasma shells, see e.g. \cite{Rees:1994nw, Daigne:1998xc, Kobayashi:1997jk} -- meaning that in that sense the assumption (simplification) of a single $R_C$ contradicts the idea of (efficient) energy dissipation in internal shocks. As a consequence,  
it has been shown that different messengers (neutrinos, gamma-rays, UHECRs) originate from different regions of the same GRB~\cite{Bustamante:2014oka,Bustamante:2016wpu}: neutrino production prefers small $R_C$ close to the photosphere (where the radiation densities are high), UHECRs prefer intermediate ``sweet spot'' $R_C$ (where the radiation densities are high enough for efficient particle accleleration, but not as high as that synchrotron losses limit the maximal energy), and very-high gamma-rays prefer large $R_C$ (where the radiation densities are low, and therefore $\tau_{\gamma \gamma}$ is low). 

\begin{figure}[t]
\begin{center}
\begin{tabular}{ccc}
\includegraphics[width=0.30\textwidth]{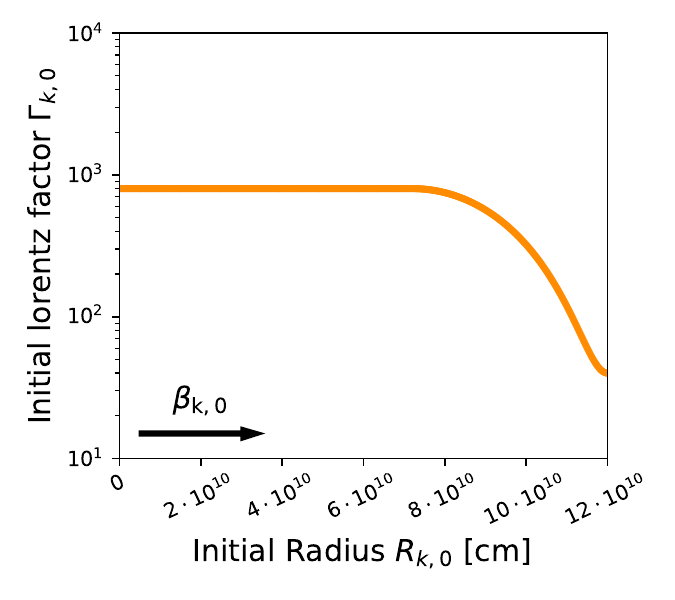} & \includegraphics[width=0.30\textwidth]{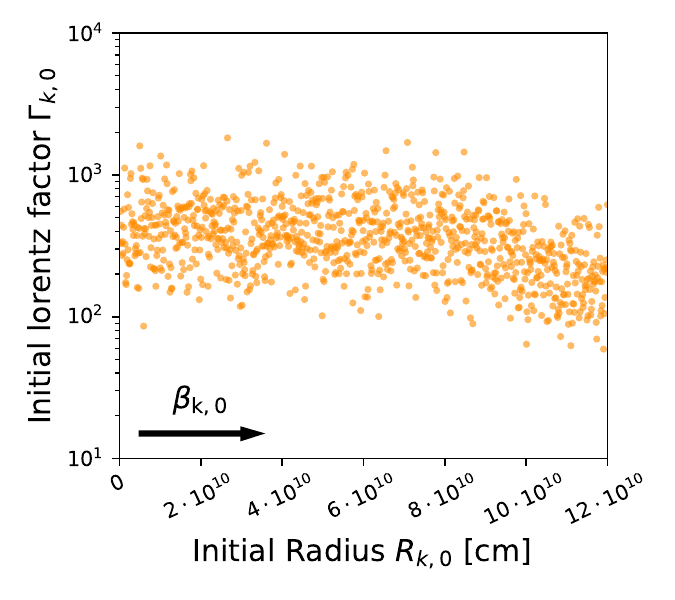} & \\[-2cm]
 \includegraphics[width=0.29\textwidth]{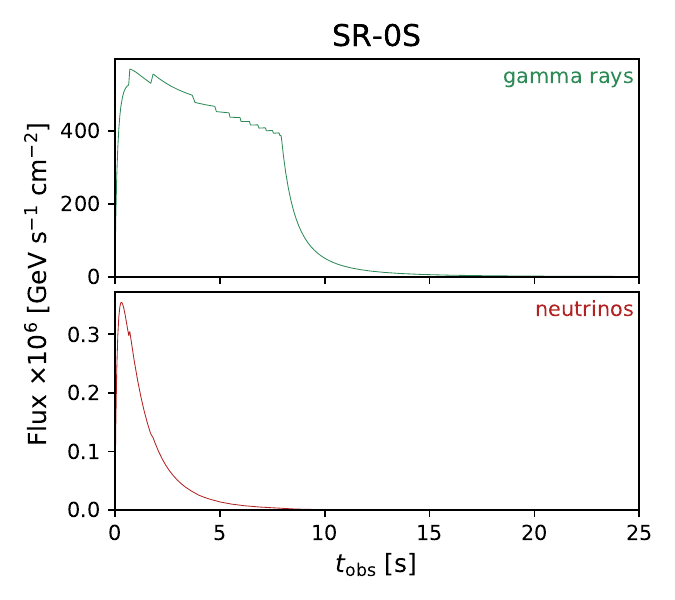} &  \includegraphics[width=0.29\textwidth]{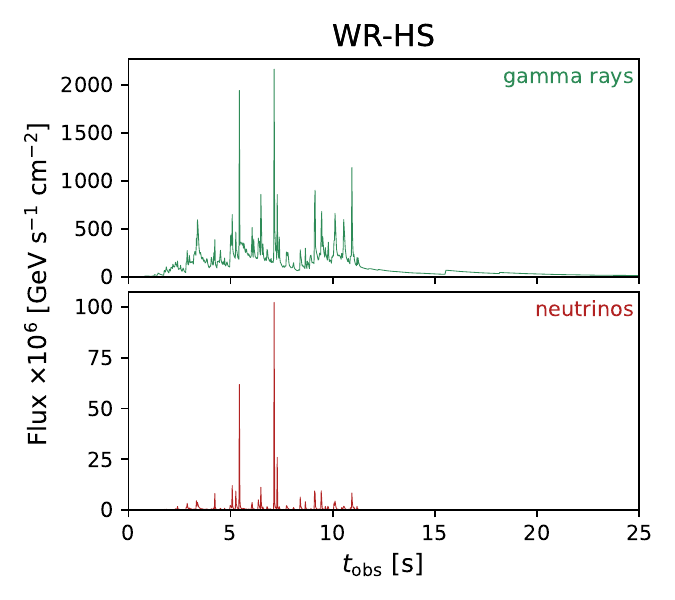} & \raisebox{+2cm}{\includegraphics[width=0.35\textwidth]{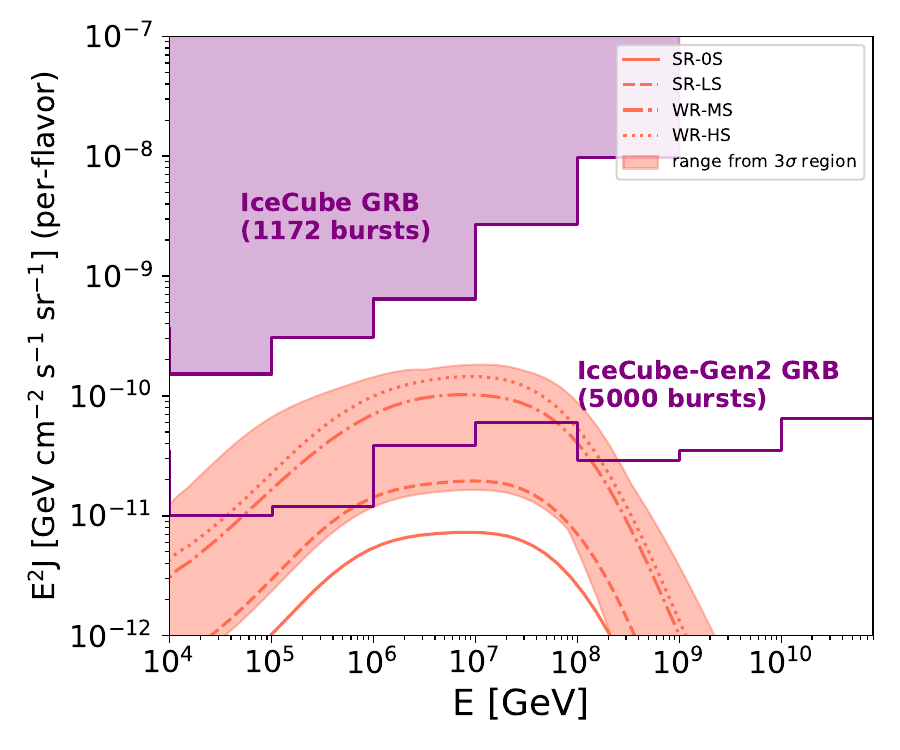}} 
\end{tabular}
\end{center}
\caption{\label{fig:grb} Left and middle columns: Two initial Lorentz factor distributions (strong engine ramp-up, no stochasticity ``SR-0S'' and weak engine ramp-up, high stochasticity ``WR-HS'') in the upper row, and the corresponding gamma-ray and neutrino light curves in the lower row. Right panel: Range of predicted neutrino fluxes for all scenarios describing UHECR data (red-shaded region) compared to different stacking limits. Figure adopted (re-arranged) from \cite{Heinze:2020zqb} (under \href{https://creativecommons.org/licenses/by/4.0/}{CC BY 4.0 DEED}).
 }
\end{figure}

It is therefore evident that the connection among the different messengers critically depends on the hydrodynamical evolution of the system, which can be traced back to the properties of the engine. We show in \figu{grb}, upper left and middle panels, two examples for the initial Lorentz factor evolution (to be read from the right to the left, as the right shells are emitted first): A strongly deterministic engine ramp-up (comparable to e.g. \cite{Bosnjak:2008bd}), and a highly stochastic ramp-up. These two examples exhibit very different gamma-ray and neutrino light curves, shown in the panels below: The left ones are smooth and the middle ones highly time-variable. The neutrino light curve in the deterministic case peaks early because there is a correlation between observation time and $R_C$, which does not exist in the stochastic case (which means that $R_C$ corresponding to the different spikes can be very different). In Ref.~\cite{Heinze:2020zqb} a general parameterization of the engine was proposed which can interpolate between these two different classes. This parameterzation was used to find the allowed parameter sets describing UHECR data (spectrum and composition), which worked reasonably well for wide parameter ranges including the ones shown here. Thus UHECR data do not seem to be very sensitive to the engine model, whereas neutrino fluences and light curves are. Most interestingly, the predicted neutrino (quasi-diffuse) flux from GRBs is shown compared to the current stacking limit in the right panel of  \figu{grb}: None of the shown models and none of the models within the $3\sigma$ fit range of UHECR data (red-shaded region) violates the neutrino stacking limit; this lower expected neutrino flux in the multi-collision model has been already pointed out in~\cite{Bustamante:2014oka}. In conclusion, the current neutrino stacking bounds are compatible with the paradigm that GRBs are the sources of the UHECRs if a more realistic (than a simple one zone) approach is used - and not even future neutrino searches can rule out this possibility completely.

What else can we learn about GRB energetics from this systematic study, especially about the GRB parameters in the context of Sec~\ref{sec:mm}? Focusing on the (exemplary) best-fit case, the gamma-ray energy has been normalized to a typical $E_\gamma^{\mathrm{iso}} \simeq 10^{53} \, \mathrm{erg}$ as an input (including about 20\% which go into sub-photospheric collisions), all other quantities are a result of the simulation and the UHECR fit: $\Gamma_{\mathrm{bulk}} \simeq 320$ and $f_e^{-1} \simeq 60$ (computed with respect to the actually emitted gamma-ray energy) within expectations -- but, interestingly, independently derived from UHECR data. The non-thermal injection composition is required to be heavy (more than 70\% energy in elements heavier than helium at the 95\% CL), which is well known problem if UHECR data are to be described. It is unclear how such a heavy composition can be achieved, one may speculate that the acceleration mechanism picks up heaver elements much more easily than light ones, see e.g. \cite{Caprioli:2017oun}.
 The required $E_{\mathrm{kin}}^{\mathrm{iso}} \simeq 4.5 \, 10^{55} \, \mathrm{erg}$ is high, of which only 13\% percent are dissipated into non-thermal radiation, mostly into cosmic-rays ($\simeq 4.6 \, 10^{54} \, \mathrm{erg}$). Since only a fraction of that energy $\simeq 4.0 \, 10^{53} \, \mathrm{erg}$ can reach the UHECR range, as discussed earlier ($f_\mathrm{bol} \simeq 12$ here) and only about 50\% can escape in that model, about   $\simeq 1.9 \, 10^{53} \, \mathrm{erg}$ are emitted as UHECRs. This number is consistent with our earlier rough estimate $10^{53} \, \mathrm{erg}$ to power UHECR data, and these values vary within about a factor of two depending on the point in parameter space. As a conclusion, there seems to be an ``energy hierarchy'' challenge $E_{\mathrm{kin}}^{\mathrm{iso}} \sim 100 \, E_{\gamma}^{\mathrm{iso}}$, which comes from a) a relatively large $f_{\mathrm{bol}}$, and b) the low dissipation efficiency of the internal shock models. While issue a) is similar to AGN blazars, and rises the question of the harder acceleration spectra or larger cosmic-ray injection energies, issue b) is a well known issue of the internal shock model (and both contribute by about a factor of ten each here). Note that although $E_{\mathrm{kin}}^{\mathrm{iso}} \simeq 4.5 \, 10^{55} \, \mathrm{erg}$ seems high from the afterglow perspective, its beaming-corrected value (for a jet opening angle $3.5^\circ$) can be comfortably extracted from the rotational energy of a rapidly spinning solar-mass-sized black hole in accreting black hole or magnetar models, see e.g. discussion in Ref.~\cite{Rudolph:2022dky}, and is also compatible with a fraction of the mass of a collapsing solar-mass star -- which means that there is no fundamental ``energy budget violation crisis'' comparable to AGN blazars.   The obtained baryonic loading $f_e^{-1} \simeq 60$ is significantly lower than the one expected from the individual neutrino observations of AGN blazars. Even energy equipartition may be achievable if a special class of energetic GRBs are the sources of the UHECRs, which may also lead to interesting hadronic secondary signatures in the electromagnetic spectrum~\cite{Asano:2008tc,Rudolph:2022ppp}. While for jetted AGN the neutrino peak energy and UHECR energies would be directly related (see Sec.~\ref{sec:mm} and \cite{Rodrigues:2020pli}), the GRB neutrino spectra typically do not exhibit this connection for strong enough magnetic fields because of the secondary (pion, muon, kaon) cooling. For larger collision radii and lower magnetic fields, however, GRB neutrinos may exhibit a similar behavior and may actually peak at higher energies -- potentially escaping the IceCube stacking searches.

\section{Tidal Disruption Events (TDEs)}

Jetted TDEs have been long time discussed as possible neutrino and UHECR sources, and also the recent neutrino-TDE associations have been studied in that context (e.g. \cite{Winter:2020ptf} for AT2019dsg). However, direct evidence for a relativistic jet in combination with a neutrino observation has not yet been established yet, and the neutrino production region of the observed neutrino-TDE associations is uncertain. It is however interesting that all of these associations have a strong dust echo in common, which (in terms of temporal evolution) matches the neutrino observation times -- which have been delayed order hundred days after peak in all cases. If these infrared photons are indeed the target photons in \equ{nuprod}, the photo-pion threshold requires UHECR primaries beyond the EeV range, which means that there is a direct connection to the origin of the UHECRs. A corresponding isotropic neutrino emission model has been proposed in \cite{Winter:2022fpf} (M-IR), where the accelerator was not specified. One possibility are jets pointing in other directions, where escaping cosmic rays are isotropized in a large interaction region filled with the infrared photons. It is interesting to observe that the fraction of neutrino-emitting TDEs and the fraction of jetted TDEs (see discussion in \cite{Piran:2023svv}) seem roughly consistent, which may be an indication that jets power the particle acceleration, but is no evidence. Note that for the neutrino production, there is a similar energetics problem as the one for AGN and GRBs: a very large fraction of the initial mass accretion rate needs to be transferred into cosmic rays; this problem  can be mitigated in models with a strongly collimated outflows, but is more challenging for all isotropic emission models~\cite{Winter:2021lyo}. We anticipate that the subject of jetted TDEs will become more attractive in this conferences series in the light of recent discoveries independent of neutrinos, but regarding the neutrino and UHECR connections especially the link to relativistic jets is somewhat speculative at this point.

\section{Summary and conclusions}

Neutrino observations in the context of relativistic outflows have so far been especially established for AGN blazars. The detection of radiation across the electromagnetic spectrum limits the properties of the production region and point towards sources with high baryonic loadings and low neutrino production efficiencies, which imposes some tension when comparing the required proton luminosity to the Eddington accretion rates; one may speculate that neutrino flares may come at times of super-Eddington accretion, while other solutions require extra ingredients, such as more compact production zones or strong external radiation fields to mitigate (but not eliminate) this challenge. 

While no neutrinos from GRBs have been seen yet, the stringent neutrino stacking limits constrain one-zone internal shock models of GRBs. If more realistic hydrodynamical outflow models with extended production regions are used, however, it cannot be ruled out that GRBs are the sources of the UHECRs. As an interesting observation in internal shock scenarios, the (isotropic-equivalent) kinetic energy of the typical GRB must exceed $10^{55} \, \mathrm{erg}$ -- which may be challenging from the afterglow perspective, but is in principle compatible with the available energy budget for moderate jet opening angles. Compared to neutrino-observed AGN, GRBs seem to exhibit lower baryonic loadings and higher neutrino production efficiencies because of more compact radiation zones.

In all cases, certain commonalities can be established from neutrino observations: a) $E^{-2}$ (or softer) acceleration spectra, as expected  from acceleration theory, are challenging from the energetics perspective as too much energy is ``bound'' at the low end of the spectrum; perhaps cosmic rays exhibit a high minimal injection energy ($E_{p,\mathrm{min}}$, similar to $E_{e,\mathrm{min}}$ frequently invoked in electromagnetic AGN models). b) Energy equipartition between electrons and protons is in most cases not compatible with neutrino data or the UHECR paradigm; typically very large baryonic loadings are required, and the overall transfer efficiency of energy into non-thermal cosmic rays must be high (which can be quantified for different object classes). c) If UHECRs are to be described, the composition of the non-thermal spectra seems to be heavier than the mass composition of the injected materials; are thus heavier ions easier accelerated?

While these observations seem obvious to the neutrino and UHECR communities, they should be interpreted as ``conditional requirements'' on the acceleration process. Some of the conclusions are certainly model-dependent, but it is interesting that most models with relativistic jets face similar challenges regarding energetics. For neutrinos, these are typically be mitigated if neutrinos come from much more compact production zones unrelated to a relativistic jet, such as AGN cores. But then the question arises: do astrophysical neutrinos and UHECRs come from different source classes?

\vspace*{0.3cm}

{\em Acknowledgments.} I would like to thank Chengchao Yuan for useful comments. This work has been supported by the European Research Council (ERC) under the European Union's Horizon
2020 research and innovation programme (Grant No. 646623).

\footnotesize{

\providecommand{\href}[2]{#2}\begingroup\raggedright\endgroup

} 

\end{document}